\documentclass[3p,times,procedia]{elsarticle}
\usepackage{custom}
\usepackage{nupha_ecrc}

\volume{00}

\firstpage{1}

\journalname{Nuclear Physics A}

\runauth{X. An}

\jid{nupha}

\jnltitlelogo{Nuclear Physics A}

\usepackage{amssymb}

\usepackage[figuresright]{rotating}

\begin{document}

\begin{frontmatter}



\dochead{XXVIIIth International Conference on Ultrarelativistic Nucleus-Nucleus Collisions\\ (Quark Matter 2019)}

\title{Relativistic Dynamics of Fluctuations and QCD Critical Point}


\author{Xin An}
\address{Department of Physics, University of Illinois, Chicago, Illinois 60607, USA}

\begin{abstract}
	We review the recent development on a systematic deterministic formalism describing dynamics of both bulk and fluctuations in an arbitrary relativistic hydrodynamic flow carrying conserved charges. In particular, we discuss the implementation of such general formalism near the QCD critical point.
\end{abstract}


\end{frontmatter}


\section{Introduction}
\label{sec:introduction}

The quark-gluon plasma created in the heavy-ion collision experiments has been manifested as a nearly perfect fluid, whose dynamic evolution is well described by hydrodynamics as the system is large enough to be treated hydrodynamically. On the other hand, the system is still small enough such that the hydrodynamic fluctuations are not negligible and are indeed essential in the vicinity of the QCD critical point. A systematic treatment of hydrodynamic fluctuations can be dated back to 1950s by Landau and Lifshitz, based on the stochastic Langevin dynamics where dealing with the divergences from the noise is inevitable. In 1970s, Andreev developed a complimentary formalism \cite{Andreev:1978} in a nonrelativistic context, by performing a renormalization procedure such that all divergences coming from the noise are absorbed into the bare hydrodynamic variables, and the resulting deterministic equations are cutoff independent and therefore simulation friendly. In recent years, this deterministic (also referred to as hydro-kinetic) approach was studied for a relativistic fluid subject to the symmetry of boost-invariance \cite{Akamatsu:2017,Akamatsu:2018,Martinez:2018}, and was partially generalized to a neutral fluid with {\em arbitrary} background in Ref.~\cite{An:2019rhf}, where a systematic formalism describing dynamics of both bulk and fluctuations was presented, yet still inadequate for its implementation near the critical point, since the conserved charges play an important role due to the critical slowing down. Extending our previous work \cite{An:2019rhf} to accommodate {\em charged} fluid and implementing it with the critical dynamics is the major goal of the work formulated in Ref.~\cite{An:2019fdc}, the main ideas and key results of which are briefly reviewed in this proceeding.

\section{General Formalism}
\label{sec:gf}

In Landau and Lifshitz's formalism, hydrodynamic fluctuations can be characterized by a set of {\em stochastic} variables (e.g., entropy density $\sm$, pressure $\sth p$ and fluid velocity $\su$) fluctuating between ensemble members of the thermal system (collision events in heavy-ion collisions), generated by the noise resulting from the coarse-graining over the hydrodynamic cells. Provided the constitutive relations $\sT^{\mu\nu}=T^{\mu\nu}(\sm,\sth p,\su)+\sS^{\mu\nu}$ and $\sJ^\mu=J^\mu(\sm,\sth p,\su)+\sI^\mu$, the hydrodynamic evolution is described by the conservation equations
\begin{equation}\label{eq:hydro-stoch}
\del_\mu \sT^{\mu\nu}=0\,,\quad
\del_\mu \sJ^\mu=0\,,
\end{equation}
where $\sT^{\mu\nu}$ and $\sJ^\mu$ are the stochastic energy-momentum tensor and charge current vector respectively, $\sS^{\mu\nu}$ and $\sI^\mu$ are the corresponding random noises whose amplitudes are set by the fluctuation-dissipation theorem, and are assumed to be statistically independent and local, i.e., $\av{\sS^{\mu\nu}(x)\sI^\lambda(x^\prime)}=0$ and $\av{\sI^\mu(x)\sI^\nu(x^\prime)}\sim\av{\sS^{\mu\nu}(x)\sS^{\lambda\kappa}(x^\prime)}\sim\delta^{(4)}(x-x^\prime)$. The $\delta$-function gives rise to the ``infinite noise'' problem -- the noise amplitude is infinitely large as the hydrodynamic cell size approaches zero, a challenge for numerical implementation. 

This awkward problem could be resolved if one switches the gear to the {\em deterministic} approach, where the ingredients one deals with are the ensemble averages of the stochastic quantities (e.g., $m\equiv \langle \sm \rangle$, $p\equiv\langle\breve p\rangle$ and $u\equiv\langle\su\rangle$) and multi-point correlation functions of fluctuations (e.g., two-point function $G_{AB}(x,y)\equiv\av{\phi_A(x+y/2)\phi_B(x-y/2)}$, where $\phi\sim(\dm,\dpp,\du_\mu)\,$ represents the properly rescaled fluctuating deviations with subscript $A,B=m,p,\mu$, and $x$ denotes the midpoint of two spacetime variables while $y$ denotes their separation). More specifically, the averaged energy-momentum tensor and charge current vector will not only depend on the aforementioned hydrodynamic variables, but also the nonlinear feedback from their fluctuations. To leading order only two-point functions are relevant and one finds
  \begin{eqnarray}\label{eq:avTJ0}
    \av{\sth T^{\mu\nu}(x)}=T^{\mu\nu}(m,p,u)+L\left[G^{mm},G^{pp},G^{m\mu},G^{p\mu},G^{\mu\nu}\right], \hspace{1.2mm} \av{\sJ^\mu(x)}=J^\mu(m,p,u)+L\left[G^{mm},G^{pp},G^{m\mu},G^{p\mu}\right]
  \end{eqnarray}
where $L\,[\dots]$ represents a linear combination of quantities in the square bracket. The bare part $T^{\mu\nu} and J^\mu$ satisfies the evolution equation for one-point functions that is well known in non-fluctuating hydrodynamics, while the fluctuation part $L\,[\dots]$ relies on a set of two-point correlation functions $G_{AB}(x)\equiv G_{AB}(x,y=0)$, whose dynamic description is in demand. One may naively expect that we would achieve this by substituting the linearized Eq.~\eqref{eq:hydro-stoch} into $u\cdot\partial\,G(x,y)$. However, dealing with a general relativistic fluid is a bit tricky: first, in the most natural choice, the concept of ``time'' is no longer global as in the lab frame but frame-dependent, thus $G_{AB}(x,y)$, as an {\em equal-time} correlator, shall be defined to comply with the condition $u(x)\cdot y=0$, where $y_\mu=y_a e_\mu^a(x)$ represents a spatial hyper-surface expanded by the triad $e_\mu^a(x)$, that is perpendicular to the local temporal vector $u(x)$; second, the variation of $G_{AB}(x,y)$ is more than a difference of which defined in different local rest frames, it also incorporates a purely kinematic change (associated with a Lorentz boost from one frame to another) that we are not interested in. Thus, we introduce the {\em confluent} derivative (connection) $\bar{\nabla}\,$ and correlator $\bar G_{AB}$ whose definition is adjusted by the fluid \cite{An:2019rhf}, such that the equal-time constraints are maintained and irrelevant degree of freedoms are eliminated. Assembled by this sophisticated manipulation, the evolution matrix equation for the confluent two-point function upon a Wigner transform turns out to be 
\begin{equation}\label{eq:kinetic_eq}
\bu\cdot\cfd \W(x,\p)= -\comm{ i \LL^{(\p)}, \W}+\dots\, \quad\text{where}\quad \W_{AB}(x,\p) \equiv \int d^4y\,\delta( u(x)\cdot y)\, e^{-i\p\cdot y}\, \GG_{AB}(x,y)
\end{equation}
is a confluent Wigner function, $[X,Y]\equiv XY-YX$ is the anti-commutator, ``$\dots$'' represents less-dominant terms in gradient expansion and are kind of cumbersome \cite{An:2019rhf}. Nevertheless, Eq.~\eqref{eq:kinetic_eq} will be significantly simplified if we diagonalize it in the eigenbasis of its dominant operator $\LL^{(\p)}$ and average out the fast oscillating modes. In terms of the rescaled function $N_{\cA\cB}\sim \psi_{\cA}^A \W_{AB} \psi_{\cB}^B$ where $\psi$ is the transformation matrix and $\cA,\cB=+,-,m,(1),(2)$ labels the eigenbasis of $\LL^{(\p)}$, the resulting equation for the decoupled longitudinal modes reads
\begin{equation}\label{eq:Npm}
\mathcal L_L[N_{++}]\equiv\left((\bu + v)\cdot\cfd-f\cdot\frac{\partial}{\partial \pp}\right) N_{++}=- \Gamma_L\left(N_{++}-N_{++}^{\text{(eq)}}\right),
\end{equation}
 which is precisely the kinetic equation for phonons! It describes the relaxation of sound mode $N_{++}$, in the rate $\Gamma_L=\gL q^2$, to its thermodynamic equilibrium $N_{++}^{\text{(eq)}}=T/E$, the equilibrium Bose-Einstein distribution of phonons with energy $E=c_s|q|$ in the high temperature limit $T\gg E$ where $c_s$ is the speed of sound. The Liouville-like operator, $\mathcal L_L$, describes the phonon propagation (in the velocity of sound $v=c_s\hat q$ where $\hat q\equiv q/|q|$) on top of an arbitrary flow background, driven by the force $f$ including relativistic inertial and Coriolis force due to acceleration and rotation of the flow, as well as the ``Hubble'' force due to the Hubble-like expansion \cite{An:2019rhf}. The remaining modes mix the fluctuations of entropy ($\cA,\cB=m$) and transverse velocities with two degree of freedoms ($\cA,\cB=(i)$ where $i=1,2$). They obey equations similar to Eq.~\eqref{eq:Npm}: 
\begin{subequations}\label{eq:LN}
\begin{align}
\mathcal{L}[N_{mm}] =& -2\Gamma_\lambda\left(N_{mm}-N_{mm}^{\text{(eq)}}\right)+L[N_{m \T{i}}]\,,\label{eq:LN_ss}\\
\mathcal{L}[N_{m \T{i}}] =& -(\Gamma_\eta+\Gamma_\lambda)\left(N_{m \T{i}}-N_{m\T{i}}^{\text{(eq)}}\right)+L[N_{mm},N_{m \T{i}},N_{\T{i}\T{j}}]\,,\label{eq:LN_si}\\
\mathcal{L}[N_{\T{i}\T{j}}] =& -2\Gamma_\eta\left(N_{\T{i}\T{j}}-N_{\T{i}\T{j}}^{\text{(eq)}}\right)+L[N_{m \T{i}},N_{\T{i}\T{j}}]\label{eq:LN_ij}\,,
\end{align}
\end{subequations}
with additional coupling terms represented by $L\,[\dots]$. $\mathcal L$ is again a Liouville-like operator, $\Gamma_\lambda=\gl q^2$ and $\Gamma_\eta=\geta q^2$ are the relaxation rates for diffusive and shear modes respectively, with which the fluctuations relax to their thermodynamic equilibrium $N_{\cA\cB}^{\text{(eq)}}$. In particular, $N_{mm}^{\text{(eq)}}\sim c_p$ where $c_p$ is the specific heat.

With our general deterministic formalism well-established, we are now in the right position to deal with the ``infinite noise'' problem. Remember, the ``infinite noise'' results from the noise correlation in thermodynamic equilibrium. However, fluctuations are driven out of equilibrium in the presence of system gradients. It is therefore convenient to decompose the two-point functions $G_{AB}$ ($N_{AB}$) into the equilibrium and non-equilibrium part: $G_{AB}(x)=G^{(\text{eq})}_{AB}(x)+G^{(\text{neq})}_{AB}(x)$. Substituting this ansatz into Eqs.~\eqref{eq:Npm}, \eqref{eq:LN} and integrating out wavenumber $q$, the divergence from ``infinite noise'' is regularized by the wavenumber cutoff $\Lambda$ (i.e., $G^{(\text{eq})}_{AB}\sim\Lambda^3$), and can be absorbed into the bare ideal constitutive relations by renormalizing relevant hydrodynamic variables. To sweep all divergences completely, we further decompose $G^{(\text{neq})}_{AB}(x)=G^{(1)}_{AB}(x)+ \wt G_{AB}(x)$, where $G^{(1)}_{AB}\sim\partial(u,m)\Lambda$ is linear in gradient and hence renormalizes the transport coefficients including viscosities and conductivity with their semi-positivity preserved, while the remaining non-analytic term $\wt G_{AB}$ is the so-called long-time tails. Labeling the renormalized quantities by ``$R$'', Eq.~\eqref{eq:avTJ0} now reads
  \begin{eqnarray}\label{eq:avTJ1}
    \av{\sth T^{\mu\nu}(x)}=T^{\mu\nu}(m_R,p_R,u_R)+L\left[\wt G^{\mu\nu}\right],~~\av{\sJ^\mu(x)}=J^\mu(m_R,p_R,u_R)+L\left[\wt G^{m\mu},\wt G^{p\mu}\right],
  \end{eqnarray} 
and the resulting cutoff-independent equations suitable for numerical simulation are (c.f. Eq.~\eqref{eq:hydro-stoch})
\begin{equation}\label{eq:conservation-TJ}
  \partial_\mu\av{\sT^{\mu\nu}(x)} =0\,, \quad \partial_\mu\av{\sJ^{\mu}(x)} =0\,.
\end{equation}
Eqs.~\eqref{eq:avTJ1}, \eqref{eq:conservation-TJ}, together with Eqs.~\eqref{eq:Npm} and \eqref{eq:LN} form a closed set of cutoff-independent, deterministic equations describing dynamics of both bulk and fluctuations.

\section{Fluctuation Dynamics Near the Critical Point}
\label{sec:cp}

In this section we outline the implementation of our general formalism near the QCD critical point. The main feature in the critical region is that, the equilibrium correlation length $\xi$, which is microscopically small away from the critical point, becomes macroscopically large in the thermodynamic limit as the system approach the critical point. In heavy-ion collisions, it could increase to be much larger than any microscopic scales, yet still limited by the size of the fireball.

Due to the nonlinear and nonlocal mode-coupling phenomena in the vicinity of the critical point, the contribution of the feedback are dominated by fluctuations at wavenumber (scale) $q\sim\xi^{-1}$, around which two modifications (approximations) shall be made upon Eqs.~\eqref{eq:LN}, i.e., 
\begin{equation}
  \label{eq:q-dependence}
  c_p \to c_p(q)=\frac{c_p}{1+(q\xi)^2}\,, \quad \gl \to \gl(q)  = \frac{\kappa(q)}{c_p(q)}\,.
\end{equation}
 The $q$-dependent specific heat $c_p(q)$ is given by the Ornstein-Zernike approximation, such that the equilibrium correlation function $G_{mm}^{(\text{eq})}(x,y)\sim e^{-|y|/\xi}/|y|$ is screened in a Yukawa form. Meanwhile, the modified relaxation coefficient $\gl(q)$ depends on $q$ also through the heat conductivity $\kappa(q)=\kappa_0+\kappa_\xi K(q\xi)$ by the Kawasaki approximation, where $\kappa_0$ and $\kappa_\xi=c_pT/6\pi\eta\xi$ are the noncritical and critical part respectively, and the latter is modified by the Kawasaki function $K(x)$ satisfying $K(0)=1$. Eqs.~\eqref{eq:LN} with the modifications given by Eq.~\eqref{eq:q-dependence} are the key equations we proposed for an extended formalism we called {\em Hydro++}.

\begin{figure}[ht]
  \centering
  \includegraphics[height=12em]{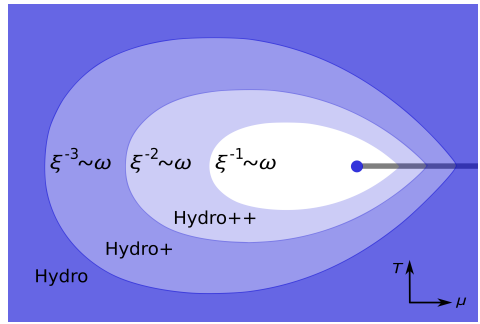}
  \caption{A schematic plot of the vicinity of the critical end point (blue point) in the $T-\mu$ plane of the QCD phase diagram. The contours of the equilibrium correlation length $\xi$ separate several distinct regimes characterized by frequency $\omega$. See text for detailed illustration.}
  \label{fig:CP}
\end{figure}

 Away from the critical point, our general formalism presented in Sec.~\ref{sec:gf} provides a systematic dynamic description. Near the critical point, the implementation of our extended formalism is more subtle, however. Due to the critical slowing down, we have $\Gamma_\lambda\sim\xi^{-3}\ll\Gamma_\eta\sim\xi^{-2}$, thus different modes in Eqs.~\eqref{eq:LN} may relax with parametrically different rates, and compete with the background evolution rate $\omega$ in different scenarios. As depicted by Fig.~\ref{fig:CP}, in the long wavelength limit $\omega\ll \Gamma_\lambda\lesssim\Gamma_\eta$, manifested away from the critical point, most wavenumber modes equilibrate rapidly compared to $\omega$, therefore ordinary hydrodynamics (Hydro) is sufficient to describe the system. As the system approaches the critical point ($\xi$ increases), $\omega$ will first fall into the window $\Gamma_\lambda\lesssim\omega\ll\Gamma_\eta$ where hydrodynamics breaks down and Hydro+ applies \cite{Stephanov:2018hydro+}, thus the slowest mode $N_{mm}$ associated with $\Gamma_\lambda$ has to be taken into account (see Eq.~\eqref{eq:LN_ss}). Hydro++, however, nontrivially extends the applicability of Hydro+ further to $\Gamma_\lambda\ll\Gamma_\eta\lesssim\omega$, i.e., closer to the critical point, such that the whole set of equations in \eqref{eq:LN} must be involved. Nonetheless, we shall emphasis that Hydro++ is limited by $\omega\ll\xi^{-1}$, since the nonlocality at scale $\omega\sim\xi^{-1}$ is not negligible and an extended formalism is still plausible.

\section{Outlook}
\label{sec:outlook}

In this essay we review the work presented in Ref.~\cite{An:2019rhf,An:2019fdc}, where the state-of-the-art formalism for deterministic fluctuating hydrodynamics, as an essential ingredient of the theoretical framework for interpreting the experimental results from the RHIC Beam Energy Scan Program, is developed. To accomplish such framework, we shall extend our formalism by including higher-point functions that are more sensitive to the critical point, and connect it to the freezeout kinetics and observables. Moreover, it would be worthwhile to consider its accommodation to the first-order phase transition line (dashed line in Fig.~\ref{fig:CP}), providing a more complete picture of fireball evolution in high baryon density region. We defer these to future work. 

~

This work is supported by the U.S. Department of Energy, Office of Science, Office of Nuclear Physics, within the framework of the BEST Topical Collaboration and grant No. DE-FG0201ER41195.





\bibliographystyle{elsarticle-num}
\bibliography{references}

\begin{thebibliography}{1}
\expandafter\ifx\csname url\endcsname\relax
  \def\url#1{\texttt{#1}}\fi
\expandafter\ifx\csname urlprefix\endcsname\relax\def\urlprefix{URL }\fi
\expandafter\ifx\csname href\endcsname\relax
  \def\href#1#2{#2} \def\path#1{#1}\fi

\bibitem{Andreev:1978}
A.~Andreev, Corrections to the hydrodynamics of liquids, Zh. Ehksp. Teor. Fiz.
  75~(3) (1978) 1132--1139.

\bibitem{Akamatsu:2017}
Y.~Akamatsu, A.~Mazeliauskas, D.~Teaney, Kinetic regime of hydrodynamic
  fluctuations and long time tails for a bjorken expansion, Phys. Rev. C 95
  (2017) 014909.
\newblock \href {http://arxiv.org/abs/1606.07742} {\path{arXiv:1606.07742}},
  \href {http://dx.doi.org/10.1103/PhysRevC.95.014909}
  {\path{doi:10.1103/PhysRevC.95.014909}}.

\bibitem{Akamatsu:2018}
Y.~Akamatsu, A.~Mazeliauskas, D.~Teaney, Bulk viscosity from hydrodynamic
  fluctuations with relativistic hydrokinetic theory, Phys. Rev. C 97 (2018)
  024902.
\newblock \href {http://arxiv.org/abs/1708.05657} {\path{arXiv:1708.05657}},
  \href {http://dx.doi.org/10.1103/PhysRevC.97.024902}
  {\path{doi:10.1103/PhysRevC.97.024902}}.

\bibitem{Martinez:2018}
M.~Martinez, T.~Sch{\"a}fer, {Stochastic hydrodynamics and long time tails of
  an expanding conformal charged fluid}, Phys. Rev. C 99~(5) (2019) 054902.
\newblock \href {http://arxiv.org/abs/1812.05279} {\path{arXiv:1812.05279}},
  \href {http://dx.doi.org/10.1103/PhysRevC.99.054902}
  {\path{doi:10.1103/PhysRevC.99.054902}}.

\bibitem{An:2019rhf}
X.~An, G.~Ba{\c s}ar, M.~Stephanov, H.-U. Yee, {Relativistic Hydrodynamic
  Fluctuations}, Phys. Rev. C 100~(2) (2019) 024910.
\newblock \href {http://arxiv.org/abs/1902.09517} {\path{arXiv:1902.09517}},
  \href {http://dx.doi.org/10.1103/PhysRevC.100.024910}
  {\path{doi:10.1103/PhysRevC.100.024910}}.

\bibitem{An:2019fdc}
X.~An, G.~Ba{\c s}ar, M.~Stephanov, H.-U. Yee, {Fluctuation dynamics in a
  relativistic fluid with a critical point. }\href
  {http://arxiv.org/abs/1912.13456} {\path{arXiv:1912.13456}}.

\bibitem{Stephanov:2018hydro+}
M.~Stephanov, Y.~Yin, Hydrodynamics with parametric slowing down and
  fluctuations near the critical point, Phys. Rev. D 98 (2018) 036006.
\newblock \href {http://arxiv.org/abs/1712.10305} {\path{arXiv:1712.10305}},
  \href {http://dx.doi.org/10.1103/PhysRevD.98.036006}
  {\path{doi:10.1103/PhysRevD.98.036006}}.

\end{thebibliography}







\end{document}